\documentclass[12pt]{article}
\usepackage{latexsym}
\usepackage{graphicx}
\newcommand{\beq}{\begin{equation}}
\newcommand{\eeq}[1]{\label{#1}\end{equation}}
\newcommand{\bea}{\begin{eqnarray}}
\newcommand{\eea}[1]{\label{#1}\end{eqnarray}}

\begin{document}
\begin{flushright}

\hfill\ \ \ {MCTP-02-01}\ \ \

{hep-th/0201062}\\
\end{flushright}




\vspace*{0.88truein}

\centerline{\bf M-theory on manifolds of $G_{2}$ holonomy: }
\vspace{0.15truein}
\centerline{\bf the first twenty years\footnote{ Talk at the {\it
Supergravity@25} conference, Stony Brook, December 2001.  }}
\vspace*{0.37truein} \centerline{\footnotesize M.  J.
DUFF\footnote{mduff@umich.edu.  Research supported in part by DOE
Grant DE-FG02-95ER40899.}} \vspace*{0.015truein}
\centerline{\footnotesize\it Michigan Center for Theoretical Physics}
\baselineskip=10pt
\centerline{\footnotesize\it Randall Laboratory, Department of Physics,
University of Michigan}
\centerline{\footnotesize\it Ann Arbor, MI 48109--1120, USA}
\bigskip
\vspace*{0.21truein}
\abstract{In 1981, covariantly constant spinors were introduced into 
Kaluza-Klein theory as a way of counting the number of supersymmetries 
surviving compactification.  These are related to the {\it holonomy} 
group of the compactifying manifold. The first non-trivial example was 
provided in 1982 by
$D=11$ supergravity on the squashed $S^{7}$, whose $G_{2}$ holonomy
yields $N=1$ in $D=4$.  In 1983 another example was provided by $D=11$
supergravity on $K3$, whose $SU(2)$ holonomy yields half the maximum
supersymmetry.  In 2002, $G_{2}$ and $K3$ manifolds continue to
feature prominently in the full $D=11$ M-theory and its dualities.  In
particular, singular $G_{2}$ compactifications can yield chiral
$(N=1,D=4)$ models with realistic gauge groups.  The notion of 
generalized holonomy is also discussed.


\setcounter{footnote}{0}
\renewcommand{\thefootnote}{\alph{footnote}}

\vspace*{-0.5pt}
\newpage
\section{Introduction}
\label{Physics}

For the {\it Supergravity@25} conference, the organizers requested that the
speakers provide a blend of something historical and something
topical.  So I have chosen to speak about {\it holonomy}, especially
$G_2$ holonomy.

In 1981, Witten laid down the criterion for spacetime 
supersymmetry in Kaluza-Klein theory \cite{Wittensearch}. The number 
of spacetime supersymmetries is given by the number of covariantly 
constant spinors on the compactifying manifold. Covariantly 
constant spinors are, in their turn, related to the {\it holonomy}  
\footnote{Holonomy had
already found its way into the physics literature via gravitational
instantons \cite{Hawkingpopesymm} and non-linear sigma models
\cite{Alvarez-Gaume}.} group of the corresponding connection. It was
well 
known that the
number of massless gauge bosons was determined by the isometry group
of the compactifying manifold, but it turned out to be the holonomy 
group that determined the number of massless gravitinos.  This was 
further emphasized in \cite{Awadaduffpope}.

The first non-trivial example was provided in 1982
\cite{Awadaduffpope} by compactifying $D=11$ supergravity on the squashed
$S^{7}$ \cite{Duffnilssonpopesuper}, an Einstein space whose whose $G_{2}$
holonomy yields $N=1$ in $D=4$.  (The round $S^{7}$ has trivial
holonomy and hence yields the maximum $N=8$ supersymmetry
\cite{DuffpopeKK}.) Although the phenomenologically desirable $N=1$
supersymmetry \cite{Wittensearch} and non-abelian gauge groups appear in four
dimensions, the resulting theory was not realistic, being vectorlike
with $SO(5) \times SO(3)$ and living on $AdS_{4}$.  It nevertheless
provided valuable insight into the workings of modern Kaluza-Klein
theories.  Twenty years later, $G_{2}$ manifolds continue to play an
important role in $D=11$ M-theory for the same $N=1$ reason.  But the
full M-theory, as opposed to its low energy limit of $D=11$
supergravity, admits the possibility of {\it singular} $G_{2}$
compactifications which can yield {\it chiral} $(N=1,D=4)$ models
living in Minkowski space and with realistic gauge groups.

In 1983 another example was provided by $D=11$ supergravity on $K3$
\cite{Duffnilssonpopecomp}, whose $SU(2)$ holonomy yields half the maximum
supersymmetry.  For the first time, the Kaluza-Klein particle spectrum
was dictated by the topology (Betti numbers and index theorems) rather
than the geometry of the compactifying manifold, which was
four-dimensional, Ricci flat and without isometries.  It was thus a
forerunner of the six-dimensional Ricci flat Calabi-Yau
compactifications \cite{Candelas}, whose $SU(3)$ holonomy yields
$(N=1,D=4)$ starting from $(N=1,D=10)$.  $K3$ compactifications also
continue to feature prominently in M-theory and its dualities.

\section{D=11 supergravity}

The low energy limit of M-theory (or $D=11$ supergravity as we used to
call it) was introduced in 1978 by Cremmer, Julia and Scherk
\cite{Cremmerjuliascherk}, not long after the discovery of supergravity
itself \cite{Ferrara,Deser}.  The unique $D=11,N=1$ supermultiplet is
comprised of a graviton $g_{MN}$, a gravitino $\Psi_M$ and $3$-form
gauge field $A_{MNP}$, where $M=0, 1, \ldots 10$, with $44$, $128$ and
$84$ physical degrees of
freedom, respectively.  Already the rank of the form, dictated by
supersymmetry to be three, presages deep connections with $2$-branes
and indeed $G_{2}$ holonomy.

The supersymmetry tranformation rule of the gravitino reduces in a
purely bosonic background to
\begin{equation}
\delta \Psi_{M}={\tilde D}_{M} \epsilon
\end{equation}
where the parameter $\epsilon$ is a 32-component anticommuting spinor,
and where
\begin{equation}
{\tilde
D}_{M}=D_{M}-
\frac{i}{144}\left(\Gamma^{NPQR}{}_{M}+8\Gamma^{PQR}
\delta^{N}{}_{M}\right)F_{NPQR}•
\label{supercovariant}
\end{equation}
Here $D_{M}$ is the usual Riemannian covariant derivative involving the usual
Levi-Civita connection $\omega_{M}$
\begin{equation}
 D_{M}=\partial_{M}-\frac{1}{4}\omega_{M}{}^{AB}\Gamma_{AB},
\end{equation}
$\Gamma^{A}$ are the $D=11$ Dirac matrices and $F=dA$.

The bosonic field equations are
\begin{equation}
R_{MN}-\frac{1}{2}g_{MN}R=
\frac{1}{3}\left(F_{MPQR}F_{N}{}^{PQR}-\frac{1}{8}g_{MN}F^{PQRS}F_{PQRS}\right)
\end{equation}
and
\begin{equation}
d*F+F \wedge F=0
\end{equation}

Being at Stonybrook, I should not forget the paper that I wrote with
Peter van Nieuwenhuizen \cite{DuffvanN} pointing out that the
4-form field strength can thus give rise to a cosmological constant
$\Lambda=-12m^{2}$ in the four dimensional subspace:
\begin{equation}
F_{\mu\nu\rho\sigma}=3m\epsilon_{\mu\nu\rho\sigma}
\end{equation}
where $\mu=0,1,2,3$ and $m$ is a constant with the dimensions of mass.
A similar conclusion was reached independently by Aurilia et al \cite{Aurilia}.
This device was then used by Freund and
Rubin \cite{Freundrubin} to effect a spontaneous compactification from $D=11$
to $D=4$, yielding the product of a four-dimensional spacetime with negative
curvature
\begin{equation}
R_{\mu\nu}=-12m^{2}g_{\mu\nu}
\end{equation}
and a seven-dimensional internal space of positive curvature
\begin{equation}
R_{mn}=6m^{2}g_{mn}
\end{equation}
where $m=1,2,\ldots 7$.  Accordingly, the supercovariant derivative also splits
as
\begin{equation}
{\tilde D}_{\mu}= D_{\mu}+m\gamma_{\mu}\gamma_{5}
\end{equation}
and
\begin{equation}
{\tilde D}_{m}= D_{m}-\frac{1}{2}m\Gamma_{m}
\end{equation}
If we choose the spacetime to be maximally symmetric but leave the internal
space
$X^{7}$ arbitrary, we are led to the $D=11$ geometry $AdS_{4} \times X^{7}$.
 The first example was provided by the choice $X^{7}=$ round $S^{7}$
 \cite{DuffpopeKK,Duffnilssonpopekaluza} which is maximally
 supersymmetric\footnote{The first Ricci flat ($m=0$) example of a
 compactification of $D=11$ supergravity was provided by the choice
 $X^{7}=T^{7}$ \cite{Cremmerjulia} which is also maximally
 supersymmetric.}.  The next example was the round $S^{7}$ with
 parallelizing torsion \cite{Englert} which preserves no supersymmetry.
 However, it was also of interest to look for something in between, and
 this is where holonomy comes to the fore.

\section{Killing spinors, holonomy and supersymmetry}

The number of supersymmetries surviving compactification depends on the 
number of covariantly constant spinors \cite{Wittensearch}. To see 
this, we look for vacuum solutions of the 
field equations for which the
the gravitino field $\Psi$ vanishes.  In order that the vacuum be
supersymmetric, therefore, it is necessary that the gravitino remain
zero when we perform a supersymmety transformation and hence that the
background supports spinors $\epsilon$ satisfying 
\begin{equation}
{\tilde D}_{M}\epsilon=0
\end{equation}
In the case of a product manifold, this reduces to
\begin{equation}
{\tilde D}_{\mu}\epsilon(x) = 0
\end{equation}
and
\begin{equation}
{\tilde D}_{m}\eta(y)= 0
\label{Killing}
\end{equation}
where $\epsilon(x)$ is a 4-component anticommuting spinor and
$\eta(y)$ is an 8-component commuting spinor.  The first equation is satisfied
automatically with our choice of $AdS_{4}$ spacetime and hence the number
of $D=4$ supersymmetries, $0\leq N \leq 8$, devolves upon the number of {\it
Killing spinors} on $X^{7}$.  They satisfy the integrability condition
\begin{equation}
[{\tilde D}_{m}, {\tilde D}_{n}] \eta= -\frac{1}{4}C_{mn}{}^{ab}\Gamma_{ab}\eta=0
\label{integrability}
\end{equation}
where $C_{mn}{}^{ab}$ is the Weyl tensor.

The subgroup of $Spin(7)$
(the double cover of the tangent space group $SO(7)$) generated by this linear
combination of $Spin(7)$ generators $\Gamma_{ab}$ corresponds to the ${\it
holonomy}$ group ${\cal H}$ \cite{Awadaduffpope}.  The number of
supersymmetries, $N$, is then given by the number of singlets
appearing in the decomposition of the $8$ of $Spin(7)$ under ${\cal
H}$ \cite{Freund}.  Some examples are given in Table \ref{holonomy}.
\begin{table}
$$\matrix{{\cal H} & 8 \rightarrow & N \cr
                ~~ & ~~ & ~~\cr
                Spin(7) & 8 & 0\cr
                  G_{2} & 1+7 & 1\cr
                  SU(3) & 1+1+3+ \bar{3}& 2\cr
                  SU(2) & 1+1+1+1+2+ \bar{2} & 4\cr
                {\bf 1} & 1+1+1+1+1+1+1+1 & 8 \cr} $$
\caption{Examples of holonomy groups and the resulting supersymmetry}
\label{holonomy}
\end{table}

Strictly speaking, the Weyl tensor characterizes the {\it restricted}
holonomy group of ${\tilde D}_{m}$.  If the space is not simply
connected there may be further global obstructions to the existence of
unbroken supersymmetries.  For example, solutions of the form
$T^{7}/\Gamma$ and $S^{7}/\Gamma$, where $\Gamma$ is a discrete group,
have vanishing Weyl tensor but admit fewer than $8$ Killing spinors
\cite{Duffnilssonpopecomp}.

The phenomenological desirability of having just one Killing
spinor, and hence just one four-dimensional supersymmetry, is
also discussed in \cite{Wittensearch}.

\section{$G_{2}$}
\label{G2}
We see that the exceptional group $G_{2}$ is of particular interest
since it yields just $N=1$ supersymmetry.  In fact, the first example
of a Kaluza-Klein compactification with non-trivial holonomy was
provided by the squashed $S^{7}$ which does indeed have ${\cal
H}=G_{2}$ \cite{Awadaduffpope,Duffnilssonpopesuper}.

 This is probably a good time to say a word about terminology.  When
 talking of the {\it holonomy} of a manifold, some authors take the word
 to refer exclusively to the Levi-Civita connection appearing in the
 Riemannian covariant derivative $D_{m}$.  According to this
 definition, the holonomy of the squashed $S^{7}$ would be $Spin(7)$.
 The group $G_{2}$ would then correspond to what mathematicians call
 {\it weak holonomy} \cite{Gray,manyach}.  A 7-dimensional Einstein
 manifold with $R_{mn}=6m^{2}g_{mn}$ has weak holonomy $G_{2}$ if it
 admits a 3-form $A$ obeying
 \begin{equation}
 dA=4m *A
 \label{three}
 \end{equation}
That such a 3-form exists on the squashed $S^{7}$ can be proved
 by invoking the single (constant) Killing spinor $\eta$.
 The required 3-form is then given by \cite{Duffnilssonpopesuper}
\begin{equation}
 A_{mnp} \sim {\bar{\eta}}\Gamma_{mnp}\eta
 \label{3-form}
\end{equation}

However, I prefer not to adopt this terminology for two reasons.
First, from a strictly mathematical point of view, one should speak
not of the holonomy of a manifold but rather of the connection on the
manifold.  Different connections on the same manifold can have
different holonomies.  Secondly, from a physical point of view, the
whole reason for being interested in holonomy in the first place is
because of supersymmetry, and here the relevant connection is not the
Levi-Cvita connection $\omega$ appearing in $D_{m}$ but rather the
generalized connection appearing in
\begin{equation}
{\tilde D}_{m}= \partial_{m}-\frac{1}{4}\omega_{m}{}^{ab}\Gamma_{ab}
-\frac{1}{2}me_{m}{}^{a}\Gamma_{a}
\label{covariant}
\end{equation}
So in the context of $M$-theory, when I speak loosely of the
holonomy of a manifold, it is the supersymmetric connection that I
have in mind \footnote{Some early papers used the term {\it Weyl
holonomy} which should probably now be abandoned.}.

Owing to this generalized connection, vacua with $m\neq 0$ present subtleties
and novelties not present in the $m=0$ case \cite{vanNwarner}, for
example the
phenomenon of {\it skew-whiffing}
\cite{Duffnilssonpopesuper,Duffnilssonpopekaluza}.  For each
Freund-Rubin compactification, one may obtain another by reversing the
orientation of $X^{7}$.  The two may be distinguished by the labels
{\it left} and {\it right}.  An equivalent way to obtain such vacua is
to keep the orientation fixed but to make the replacement
$m\rightarrow -m$.  So the covariant derivative (\ref{covariant}), and
hence the condition for a Killing spinor, changes but the
integrability condition (\ref{integrability}) remains the same.  With
the exception of the round $S^{7}$, where both orientations give
$N=8$, at most one orientation can have $N \geq 0$.  This is the {\it
skew-whiffing theorem}.  A corollary is that other {\it symmetric
spaces}, which necessarily admit an orientation-reversing isometry,
can have no supersymmeties.  Examples are provided by products of
round spheres.  Of course, skew-whiffing is not the only way to obtain
vacua with less than maximal supersymmetry.  A summary of known
$X^{7}$, their supersymmetries and stability properties is given in
\cite{Duffnilssonpopekaluza}.  Note, however, that skew-whiffed vacua
are automatically stable at the classical level since skew-whiffing
affects only the spin $3/2$, $1/2$ and $0^{-}$ towers in the
Kaluza-Klein spectrum, whereas the criterion for classical stability
involves only the $0^{+}$ tower \cite{DNP,Duffnilssonpopekaluza}.

Once again the squashed $S^{7}$ provided the
first non-trivial example: the left squashed $S^{7}$ has $N=1$ but the right
squashed $S^{7}$ has $N=0$.  Interestingly enough, this means that
setting the suitably normalized 3-form (\ref{3-form}) equal to the $D=11$
supergravity 3-form provides a solution to the field equations, but
only in the right squashed case.  This solution is called the {\it
right squashed $S^{7}$ with torsion} \cite{Duffnilssonpopesuper} since
$A_{mnp}$ may be interpreted as a Ricci-flattening torsion.  Other
examples were provided by the left squashed $N(1,1)$ spaces
\cite{Page}, one of which has $N=3$ and the other $N=1$, while the
right squashed counterparts both have $N=0$.

All this presents a dilemma.  If the Killing spinor
condition changes but the integrability condition does not, how does
one give a holonomic interpretation to the different supersymmetries?
Indeed $N=3$ is not allowed by the usual rules.  The answer to this
question may be found in a paper \cite{Castellani} written before we
knew about skew-whiffing.  The authors note that in (\ref{covariant}),
the $SO(7)$ generators $\Gamma_{ab}$, augmented by presence of
$\Gamma_{a}$, together close on $SO(8)$.  Hence one may introduce a
generalized holonomy group ${\cal H}_{gen}\subset SO(8)$ and ask how
the $8$ of $SO(8)$ decomposes under ${\cal H}_{gen}$.  In the case of
the left squashed $S^{7}$, ${\cal H}_{gen}= SO(7)^{-}$, $8 \rightarrow
1+7$ and $N=1$, but for the right squashed $S^{7}$, ${\cal H}_{gen}=
SO(7)^{+}$, $8 \rightarrow 8$ and $N=0$.

Kaluza-Klein compactification of supergravity was an active area of
research in the early 1980s.  Some early papers are
\cite{Wittensearch,Awadaduffpope,Duffnilssonpopesuper,DuffpopeKK,Duffnilssonpopecomp,
Englert,Castellani,Sezgin,Freedman,Bais,Warner} and I am glad to see
many of the pioneers in the audience today: Leonardo Castellani,
Bernard de Wit, Pietro Fre, Dan Freedman, Gary Gibbons, Bernard Julia,
Ergin Sezgin, John Schwarz, Peter van Nieuwenhuizen, Nick Warner and
Peter West.  Reviews of Kaluza-Klein supergravity may be found in
\cite{Duffnilssonpopekaluza,CDF}.

\section{Supermembranes with fewer supersymmetries}

Interest in $AdS_{4}\times X^{7}$ solutions of $D=11$ supergravity
waned for a while but was revived by the arrival of the $D=11$
supermembrane \cite{Bergshoeff1}.  In 1991 this 2-brane was recovered
as a solution of the $D=11$ supergravity theory preserving one half of
the supersymmetry \cite{Duffstelle,Khuristring}.  Specifically, in the
case that $N$ branes with the same charge are stacked together, the
metric is given by
\begin{equation}
ds^2=(1+Na^{6}/y^6)^{-2/3}dx^{\mu}dx_{\mu}+
(1+Na^{6}/y^6)^{1/3}(dy^2+y^2d\Omega_7{}^2)
\label{branesolution}
\end{equation}
and the four-form field strengths by
\begin{equation}
{\tilde F}_{7}\equiv *F_4=\pm 6Na^{6}\epsilon_7
\end{equation}
Here $d\Omega_7{}^2$ corresponds to the round $S^{7}$.

Of particular interest is the near horizon limit
$y\rightarrow 0$, or equivalently the large $N$ limit, because then
the metric reduces to \cite{GT,DGT,GHT} the $AdS_{4}\times S^{7}$
vacuum with
\begin{equation}
m^{-6}=Na^{6}
\end{equation}
Thus
\begin{equation}
ds^{2}={y^{4}}{m^{4}}dx^{\mu}dx_{\mu}+{m^{-2}}{y^{-2}}dy^{2}
+m^{-2}d\Omega_{7}^{2}
\end{equation}
which is just $AdS_{4}\times S^{7}$ with the $AdS$ metric written in
horospherical coordinates.

Note that the round $S^{7}$ makes its appearance.  The question naturally
arises as to whether the compactifications with fewer supersymmetries
discussed above also arise as near-horizon geometries of $p$-brane
solitons.  The answer is yes and the soliton solutions are easy to
construct \cite{DLPS,ccdfft}.  One simply makes the replacement
\begin{equation}
d\Omega_{7}{}^{2} \rightarrow d{\hat \Omega}_{7}{}^{2}
\end{equation}
in (\ref{branesolution}), where $d{\hat \Omega}_{7}{}^{2}$ is the
metric on an arbitrary Einstein space $X^{7}$ with the same scalar
curvature as the round $S^{7}$.  The space need only be Einstein; it
need not be homogeneous \cite{DLPS}.  (There also exist brane
solutions on Ricci flat $X^{7}$ \cite{DLPS}).  Note, however, that
these non-round-spherical solutions do not tend to Minkowski
space as $r\rightarrow \infty$.  Instead the metric on the
$8$-dimensional space transverse to the brane is asymptotic to a
generalized cone
\begin{equation}
ds_{8}{}^{2}=dy^{2}+y^{2}d{\hat \Omega}_{7}{}^{2}
\label{cone}
\end{equation}
and $8$-dimensional translational invariance is absent except when
$X^{7}$ is the round $S^{7}$.  Note, however that the solutions have
no conical singularity at $y=0$ since the metric tends to $AdS_{4}
\times X^{7}$.  By introducing the Schwarzschild-like coordinate $r$
given by
\begin{equation}
r^{6}=y^{6}+Na^{6}
\end{equation}
we can see that the solutions exhibits an event-horizon at $r=N^{1/6}a$.
Indeed the solution may be analytically continued down to $r=0$ where
there is a curvature singularity, albeit hidden by the event horizon
\cite{DGT}.

As a matter of fact, there is a one-to-one correspondence between
Killing spinors on Einstein manifolds $X^{7}$ satisfying ${\tilde
D}_{m} \eta=0$, $m=1,2\ldots 7$, and Killing spinors on the Ricci-flat cone
(\ref{cone}) satisfying $D_{M}\eta=0$, $M=1,2\ldots 8$, \cite{Bar,manyach}.
So $N=1$ can then be understood as $G_{2}$ holonomy on $X^{7}$ or
$Spin(7)$ holonomy on the cone.  Similarly, the weak holonomy 3-form
(\ref{three}) lifts to a covariantly constant self-dual $4$-form on
the cone.

\section{A speculation on Spin(9)}
\label{spin9}
In Berger's classification \cite{Berger} of holonomy groups of Levi-Civita
connections given in Table \ref{Berger}, there are three exceptional cases
$G_{2}$, $Spin(7)$ and $Spin (9)$.  $G_{2}$ and $Spin(7)$ have already made
their appearance in our story, but $Spin(9)$ seems an unlikely
candidate since it corresponds to the holonomy of a {\it sixteen}-dimensional
space (the Cayley plane), and that seems too high for $D=11$
supergravity.  
\begin{table}
$$\matrix{{\cal H} & Space & restrictions \cr ~~ & ~~ & ~~\cr SO(n) &
R^{n}• & n \geq 2 \cr U(m) & R^{n}\simeq C^{m}& m \geq 2, n=2m \cr
SU(m) & R^{n}\simeq C^{m}& m \geq 2, n=2m\cr Sp(1) \times Sp(m) &
R^{n} \simeq H^{m} & m \geq 1, n=4m \cr Sp(m)& R^{n} \simeq H^{m }& m
\geq 1, n=4m\cr G_{2}&R^{7}&~~\cr Spin(7)& R^{8}&~~\cr
Spin(9)& R^{16}&~~\cr} $$
\caption{Holonomies of Riemannian connections}
\label{Berger}
\end{table}
Nevertheless, as we shall now describe, there is a way in which
subgroups of 
$SO(16)$ can appear as holonomies in the theory.  (The Cayley Plane,
which 
is the $16$-dimensional coset $F_{4}/Spin(9)$, has also featured in 
recent, but apparently unrelated, speculations on hidden mathematical 
structures in $D=11$ \cite{Ramond}.  Recent discussions on the
mathematics of $Spin (9)$ 
may be found in \cite{Friedrich1,Friedrich2}.)

The supermembranes discussed in the previous section preserve a fraction
$\nu=N/16$ of the spacetime supersymmetry where $1\leq N\leq 8$ is the
number of Killing spinors on $X^{7}$.  (In the near-horizon limit this
doubles.) Following \cite {Duffstelle}, we can attempt to quantify
this in terms of a holonomy even more generalized than that discussed
in section (\ref{G2}), namely
\begin{equation}
{\cal H}_{gen} \subset SO(16)
\end{equation}
where $N$ is then given by the number of singlets appearing in the
decomposition of the $16$ of $SO(16)$ under ${\cal H}_{gen}$.  In the
case of the brane with a round $S^{7}$, we have ${\cal H}_{gen}=SO(8)$,
the $16$ 
decomposes into an $8$ plus $8$ singlets and $\nu=1/2$.  Whereas in the
case 
of the brane with a squashed $S^{7}$, we would have only one singlet 
and $\nu=1/16$.  The origin of this $SO(16)$ was speculated in 
\cite{Fradkin} and proved in \cite{Nicolai}.  After making a $D=11$ 
Lorentz-non-covariant three/eight split of the $D=11$ supergravity 
field equations, the tangent space group $SO(1,2) \times SO(8)$ gets 
enlarged to an $SO(1,2) \times SO(16)$ under which the supersymmetry 
parameter transforms as a $(2,16)$.  Note that the $16$ is the {\it 
vector} representation, even though $\epsilon$ is a spacetime spinor 
\footnote{This phenomenon had been noted long ago by Cremmer and Julia 
\cite{Cremmerjulia} when {\it dimensionally reducing} $D=11$ 
supergravity (or Type IIB supergravity) to $D=3$, but here we are 
claiming $SO(16)$ already in $D=11$ (or $D=10$).}.  This can be 
understood by noting, as we have already done, that the supercovariant 
derivative (\ref{supercovariant}) involves not merely the Levi-Civita 
connection but extra terms involving the $4$-form field strength 
$F_{MNPQ}$, a fact that tends to be underemphasized in recent works on 
holonomy and $D=11$ supergravity (a notable exception being 
\cite{Ali}).

Although we do not know of any examples, it is thus possible that some
$D=11$ field configuration could have
\begin{equation}
Spin(9)={\cal H}_{gen} \subset SO(16)
\end{equation}
However, the embedding is such that $16 \rightarrow 16$ and so such a
configuration can preserve no supersymmetries, which may seem
something of an anticlimax.

All this raises an interesting question: What is the number of possible 
supersymmetries allowed in M-theory?  A priori, from the M-theory
algebra 
\cite{Townsend9}, $32\nu$ can be any integer from $0$ to $32$
\cite{Gauntletthull1}.  
Although there are supersymmetric backgrounds realizing the 
maximum $32$, it was thought for a time that partially supersymmetric 
backgrounds were restricted to $0 \leq \nu \leq 1/2$, such as the 
above mentioned membranes.  
However, recent work on pp waves \cite{Penrose,Michelson,Cvetic:2002si, 
Gauntletthull2, Michelson2} and G\"{o}del universes
\cite{Gauntlett:2002nw} 
has revealed certain values of $32\nu$ lying between $16$ and $32$ .  
Riemannian holonomy cannot account for these exotic fractions, so let 
us therefore take the generalized holonomy approach, discussed above.  
Of course, one is not obliged to make a three/eight split of M-theory, 
one could make any $d/(11-d)$ split.  For example, after a four/seven 
split, the tangent space group $SO(1,3) \times SO(7)$ is enlarged to 
an $SO(1,3) \times SU(8)$ \cite{Fradkin,Dewitnicolai2}.  Similar 
remarks apply to Type IIB.  It is curious to note that if the $32$ 
supercharges always belong to representations such as $(2,16)$, 
$(4,8)$ etc, then $n=32\nu$, the number of singlets appearing in the 
decomposition, is restricted to 
$0,1,2,3,4,5,6,8,10,12,14,16,18,20,22,24,26,28,32$.  This is 
consistent with the data. However, a better understanding of the two/nine and one/ten 
splits is necessary before ruling out other values.

\section{Hopf dualities}

In recent times, both perturbative and non-perturbative effects of 
ten-dimensional superstring theory have been subsumed by an 
eleven-dimensional $M$ theory whose low-energy limit is $D=11$ 
supergravity \cite{World}.  In particular, we have the duality
\begin{equation}
M~on~X \equiv IIA~on~Y
\label{duality}
\end{equation}
In the first examples $X$ was just a direct product $Y \times S^{1}$
and in this way one could give an $D=11$ M-theory origin to $D=10$
Type $IIA$ objects by either wrapping around the $S^{1}$ or by reducing.
 For
example, the $IIA$ string comes from the $M2$-brane by wrapping
\cite{Howe,Townsendeleven,Wittenvarious}; the $D6$-brane in $D=10$
may be interpreted as a Kaluza-Klein monopole in $D=11$
\cite{Dufflublack,Townsendeleven} and the $D2$ brane in $D=10$ may be
interpreted as an $M2$-brane in $D=11$ by dualizing a vector into a
scalar on the $3$-dimensional worldvolume \cite{Luscan,TownsendM}.

However, the duality (\ref{duality}) may be generalized to the so-called
{\it Hopf duality} where $X$ is a twisted product or $U(1)$ bundle over
$Y$ \cite{SWS, DLP1,DLP2,Halyo,MT}.  In \cite{SWS}, for example, such 
$M$-theory vacua with $N>0$ supersymmetry were presented which, from 
the perspective of perturbative Type $IIA$ string theory, have $N=0$.  
They can emerge whenever the $X^{7}$ is a $U(1)$ bundle over a 
$6$-manifold $Y^{6}$.  The missing superpartners are Dirichlet 
$0$-branes.  Someone unable to detect Ramond-Ramond charge would thus 
conclude that these worlds have no unbroken supersymmetry.  In 
particular, the gravitinos (and also some of the gauge bosons) are 
$0$-branes not seen in perturbation theory but which curiously remain 
massless however weak the string coupling.  The simplest example of 
this phenomenon is provided by the maximally-symmetric $S^7$.  
Considered as a compactification of $D=11$ supergravity, the round 
$S^7$ yields a four dimensional $AdS$ spacetime with $N=8$ 
supersymmetry and $SO(8)$ gauge symmetry, for either orientation of 
$S^7$.  The Kaluza-Klein mass spectrum therefore falls into $SO(8)$ 
$N=8$ supermultiplets.  In particular, the massless sector is 
described \cite{Duffnilssonpopekaluza} by gauged $N=8$ supergravity 
\cite{Dewitnicolai}.  Since $S^7$ is a $U(1)$ bundle over $CP^3$ the 
same field configuration is also a solution of $D=10$ Type $IIA$ 
supergravity \cite{Nilssonpope}.  However, the resulting vacuum has 
only $SU(4) \times U(1)$ symmetry and either $N=6$ or $N=0$ 
supersymmetry depending on the orientation of the $S^7$.  The reason 
for the discrepancy is that the modes charged under the $U(1)$ are 
associated with the Kaluza-Klein reduction from $D=11$ to $D=10$ and 
are hence absent from the Type $IIA$ spectrum originating from the 
massless Type $IIA$ supergravity.  In other words, they are Dirichlet 
$0$-branes \cite{Polchinski} and hence absent from the perturbative 
string spectrum.  There is thus more non-perturbative gauge symmetry 
and supersymmetry than perturbative.  (Here the words ``perturbative'' 
and ``non-perturbative'' are shorthand for ``with and without the 
inclusion of Dirichlet $0$-branes'', but note that the Type $IIA$ 
compactification has non-perturbative features even without the 
$0$-branes \cite{SWS}).  The right-handed orientation is especially 
interesting because the perturbative theory has no supersymmetry at 
all!  A summary of perturbative versus non-perturbative symmetries is 
given in Table 2.  In particular, the non-perturbative vacuum may have 
unbroken supersymmetry even when the perturbative vacuum has none.  
\begin{center} \begin{tabular}{|c|cc|cc|}\hline Compactification& 
&Perturbative Type $IIA$& &Nonperturbative M-theory\\ \hline\hline
Left round $S^7$ & $N=6$ & $SU(4) \times U(1)$ & $N=8$ & $SO(8)$\\
Right round $S^7$ & $N=0$ & $SU(4) \times U(1)$ & $N=8$ & $SO(8)$\\
Left squashed $S^7$ & $N=1$ & $SO(5) \times U(1)$ & $N=1$ &  $SO(5)
\times SU(2)$\\ Right squashed $S^7$ & $N=0$ & $SO(5) \times U(1)$ &
$N=0$ &  $SO(5) \times SU(2)$\\Left $M(3,2)$ & $N=0$ & $SU(3) \times
SU(2) \times U(1)$ & $N=2$ & $SU(3) \times SU(2) \times U(1)$\\
Right $M(3,2)$ & $N=0$ & $SU(3) \times SU(2) \times U(1)$ & $N=0$ &
$SU(3) \times SU(2) \times U(1)$\\ \hline
\end{tabular}\end{center}
\bigskip
\centerline{Table 2: Perturbative versus non-perturbative symmetries.}
\label{symmetries}
\bigskip
We have focussed in this paper on compactifications from $D=11$ but much of
the discussion applies, {\it mutatis mutandis } to Type $IIB$.  For
example we have the Hopf T-duality
\begin{equation}
IIB~on~S^{5}• \equiv IIA~on~CP^{2} \times S^{1}•
\label{duality1}
\end{equation}
which untwists the $S^{5}$ \cite{DLP1}.

\section{Recent developments}

Following the M-theory revolution of 1995, it was noted
\cite{Cadavid,Papadopoulos,Schwarz3,Lowe,%
Chaudhuri,Acharya,Aspinwall4} that non-chiral $N=1$ heterotic string
compactifications can be dual to $D=11$ supergravity compactified on
Ricci-flat seven-dimensional spaces of $G_2$ holonomy
\cite{joyce,joyce2,joyce3,gibbons,bryant}.  See also
\cite{Becker,Hawking} for $N=1$ compactifications on $8$-manifolds of
spin(7) holonomy.

In 2002, $G_{2}$ manifolds continue play an important role in $D=11$
M-theory for the same $N=1$ reason as in 1982 . Of course, for phenomenology
we require not only $N=1$ but also chirality, and the lack of chirality
on smooth seven-manifolds \cite{Wittenchiral} was one of the main
reasons that $D=11$ supergravity fell out of favor.  But the full
M-theory, as opposed to its low energy limit, admits the possibility
of {\it singular} $G_{2}$ compactifications which can yield {\it
chiral} $(N=1,D=4)$ models living in Minkowski space, with realistic
gauge groups \cite{acharya,amv,aw,csq,witt,%
achtwo,agva,otheracha,gomis,partouche,edelstein,kachru,%
part,kon,brand,brand2,cvglp,cmany,cvmore,stelle,achw,%
Cvetic1,harvey,bilal} 
and doublet-triplet splitting \cite{Wittensplit}.
 
\section{K3}

In 1983 a second example of the utility of holonomy was provided by
the compactification of $D=11$ supergravity on the $K3$
\cite{Duffnilssonpopecomp}.  This manifold had recently made its appearance
in the physics literature as a gravitational instanton
\cite{Hawkingpopesymm}.  It is a self-dual, and hence Ricci flat, 
solution with $m=0$ and ${\cal H}=SU(2)$ instead of the $SU(2) \times 
SU(2)$ of a generic 4-manifold.  Hence $N=N_{max}/2$.  $K3$ provided a 
novel phenomenon in Kaluza-Klein theory: the appearance of massless 
particles as a consequence of the {\it topology}, as opposed to the 
{\it geometry} of the compactifying manifold, determined by Betti 
numbers and index theorems \cite{Duffnilssonpopecomp}.  A discussion 
of boson and fermion zero modes on $K3$ and their relation to axial 
and conformal anomalies \cite{CD1}, may be found in 
\cite{Hawkingpopesymm,CD2,Wittensearch}.

$K3$ was thus the forerunner of the very influential {\it Calabi-Yau}
compactifications\footnote{Indeed, the possibility of going from 10
dimensions to 4 on a Ricci-flat 6-manifold with $SU(3)$ holonomy so as
to get $N=1$ in $D=4$ occurred to Pope, Nilsson and myself while
writing up the $K3$ paper in 1983.  Since we were at UT, Austin, at
the time, we consulted one or two of the distinguished mathematicians
in the Mathematics Department there, but were told they had never
heard of such a thing!  Consequently our paper states `` We do not
know any solutions with ${\cal H}=SU(3)$''.  } of ten-dimensional
supergravity and string theory \cite{Candelas} and, indeed, the
Ricci-flat $G_{2}$ compactifications of $M$-theory mentioned above,
some of which correspond to $K3$ fibrations \cite{achw}.

$K3$ compactification allows for the possibility of chirality in the
lower dimensional spacetime and chiral compactification of the Type $IIB$
supergravity was undertaken in \cite{townsendk3} and the heterotic
string theory in \cite{west}.

In 1986, it was pointed out \cite{Nilsson1}
that $D=11$ supergravity on $R^{10-n}\times K3\times T^{n-3}$
\cite{Duffnilssonpopecomp} and the $D=10$ heterotic string on
$R^{10-n}\times T^{n}$ \cite{Narain} not only have the same
supersymmetry but also the same moduli spaces of vacua, namely

\begin{equation}
{\cal M}=\frac{SO(16+n,n)}{SO(16+n)
\times SO(n)}
\label{moduli}
\end{equation}
It took almost a decade for this ``coincidence'' to be explained but we
now know that $M$-theory on $R^{10-n}\times K3\times T^{n-3}$ is dual to
the heterotic string on $R^{10-n}\times T^{n}$ \cite{Hulltownsend,
Wittenvarious}.

One way to understand this is to note that, when wrapped around $K3$
with its $19$ self-dual and $3$ anti-self-dual $2$-forms, the $d=6$
worldvolume fields of the $M5$-brane (or $D5$-brane)
$(B^-{}_{\mu\nu},\lambda^I,\phi^{[IJ]})$ reduce to the $d=2$
worldsheet fields of the heterotic string in $D=7$ (or $D=6$)
\cite{Townsenddual,Harveystrom,Schwarz}.  The $2$-form yields $(19,3)$
left and right moving bosons, the spinors yield $(0,8)$ fermions and
the scalars yield $(5,5)$ which add up to the correct worldsheet
degrees of freedom of the heterotic string.  A consistency check is
provided by the derivation of the Yang-Mills and Lorentz
Chern-Simons corrections to the Bianchi identity of the heterotic
string starting from the $M5$-brane Bianchi identity \cite{DLM}.

Heterotic strings on $K3$ can also be self-dual \cite{DM,DMW}.
Moreover, Heterotic strings on Calabi-Yau manifolds that are a $T^{3}$
fibration over a base $Q$ can be dual to M-theory on $X^{7}$ that is
$K3$ fibered over $Q$ and has $G_{2}$ holonomy \cite{achw}, which
brings us back to where we started.

\section{Conclusions}

I first wrote about supergravity in a popular article for New
Scientist in 1977 \cite{ns}, where I said `` Supergravity is
theoretically very compelling, but it has yet to prove its worth by
experiment''; a remark still unfortunately true at {\it
Supergravity@25}.  Let us hope that by the {\it Supergravity@50}
conference, or before, we can say something different.

It is a privilege to have been a member of the Supergravity community
these 25 years and I would like to say ``Thanks for the
memories'' to its discoverers.

\section{Acknowledgements}

I would like to thank the conference organizers, Martin Rocek, George
Sterman and Warren Siegel for their hospitality. I have enjoyed useful
conversations on holonomy with Pietro Fre, Finn Larsen, Jim Liu, Hong
Lu, Leo 
Pando-Zayas and Hisham Sati.  Thanks to Gordy Kane and Jim Liu for 
suggesting improvements to the manuscript.

\section{Notes added}
\begin{itemize}
\item
Generalized holonomy is developed further in \cite{Duff:2003ec,Hull:2003mf,Batrachenko:2003ng,Lu:2005im}.
\item
In section (\ref{spin9}) we conjectured, albeit on flimsy evidence, that the number of vacuum supersymmetries allowed by M-theory is restricted to
\begin{equation}
n=0,1,2,3,4,5,6,8,10,12,14,16,18,20,22,24,26,28,32
\label{list}
\end{equation}

  Interestingly enough, 
after the completion of this work, a G\"{o}del universe with $n=14$ 
was discovered \cite{Harmark} which completes this 
list. Furthermore:
\item
$n=31$ has now been ruled out for both Type IIB \cite{Gran:2006ec}
and Type IIA \cite{Bandos:2006xz}.
\item  $n=30$ has now been ruled out for M-theory \cite{Gran:2010tj}. 
\item The class of M-theory plane waves found in \cite{Gran:2010tj} has $n=16,20,26$ but not $n=28$, although plane wave solutions with $n=28$ do appear in Type IIB \cite{Bena:2002kq}.
\item
Backgrounds with $n>24$ are necessarily (locally) homogeneous. See \cite{FigueroaO'Farrill:2004mx} where it is also conjectured that 24 is the minimum number  which guarantees this.

\item
In compactifying Type II strings from $D=10$ to $D=2$ we must allow for the possibility of  asymmetric orbifolds where the left and right movers may experience different holonomies yielding $D=2$ supersymmetries $(N_+,N_-)$ with $N_+\neq N_-$.  
Berger's \cite{Berger} list of holonomies $SO(8), Spin(7), G_2, SU(3),SU(2),I$ allows $N_+=0,1,2,4,8,16$ and $N_-=0,1,2,4,8,16$ and hence $n=N_+ +N_-$ can take on values
\begin{equation}
n=0,1,2,3,4,5,6,8,9,10,12,16,17,18,20,24,32
\end{equation}
So if we broaden our definition of M-theory vacua to include asymmetric orbifold compactificataions of Type II  (as suggested to me in this context by Cumrun Vafa) then we must also include the cases $n=9,17$ excluded in (\ref{list}).
\end{itemize}

\end{document}